# Fluorinated graphene films with graphene quantum dots for electronic applications

I. V. Antonova,[1,2,a)] N. A. Nebogatikova,[1] and V. Ya. Prinz[1]
[1]*Rzhanov Institute of Semiconductor Physics, Russian Academy of Sciences, Siberian Branch, Novosibirsk 630090, Russia*
[2]*Novosibirsk State University, Novosibirsk 630090, Russia*

This work analyzes carrier transport, the relaxation of non-equilibrium charge, and the electronic structure of fluorinated graphene (FG) films with graphene quantum dots (GQDs). The FG films with GQDs were fabricated by means of chemical functionalization in an aqueous solution of hydrofluoric acid. High fluctuations of potential relief inside the FG barriers have been detected in the range of up to 200 mV. A phenomenological expression that describes the dependence of the time of non-equilibrium charge emission from GQDs on quantum confinement levels and film thickness (potential barrier parameters between GQDs) is suggested. An increase in the degree of functionalization leads to a decrease in GQD size, the removal of the GQD effect on carrier transport, and the relaxation of non-equilibrium charge. The study of the electronic properties of FG films with GQDs has revealed a unipolar resistive switching effect in the films with a relatively high degree of fluorination and a high current modulation (up to ON/OFF $\sim 10^4$–$10^5$) in transistor-like structures with a lower degree of fluorination. 2D films with GQDs are believed to have considerable potential for various electronic applications (nonvolatile memory, 2D connections with optical control and logic elements). *Published by AIP Publishing.*
[http://dx.doi.org/10.1063/1.4953239]

## I. INTRODUCTION

Opening and tuning the band gap of graphene-based materials may expand their applications in nano- and optoelectronics. The band gap engineering of graphene is an important and practical task.[1–5] Various graphene/fluorographene structures have attracted increasing attention in light of the creation of transistors and other devices due to the dependence of the fluorographene band gap on the F/C atom ratio.[3] A lateral graphene field-effect transistor with a channel consisting of graphene/fluorographene/graphene and providing a potential barrier to lateral carrier transport and an ON-OFF ratio of $10^5$ was demonstrated by Moon *et al.*[6] Transistor structures with graphene nanoribbons were created using lithography with a polymer mask at the region of sample fluorination for isolating the graphene ribbon from the rest of the structure.[7] Chemical isolation provides several advantages over structural isolation (lithography). First, it produces structure edges that are robust and chemically well-defined. These better defined edges appear to help retain carrier mobility. Second, chemical isolation within a larger graphene film significantly reduces adsorbate permeation, mostly starting at the active structure edges, which degrades its electronic performance.[7] The study of the properties of the fluorographene film as an insulator film gives such parameters as a permittivity of $\varepsilon = 1.2$ and the breakdown field strength of up to $10^7$ V/cm.[8] The low value of charge in fluorinated graphene (FG)-films and at interfaces with semiconductors ($\sim 2$–$5 \times 10^{10}$ cm$^{-2}$) was calculated from the capacitance–voltage characteristic of metal/fluorinated films/Si (GaAa, InAs) structures.[9] It was demonstrated that graphene fluorination causes band gap widening, with values of 1.8 eV and 2.9 eV for $CF_{0.10}$ and $CF_{0.48}$, respectively.[5] The height of the lateral potential barrier was estimated at 0.26 eV for the n-doped graphene/fluorographene heterostructure channel.[6] P-doped graphene quantum dots (GQDs) embedded in FG films give the values of 0.12 and 0.35 eV for potential barriers, connected with carrier activation from different quantum confinement levels in GQDs.[10] Fluorographene nanosheets are used for interface engineering in organic field-effect transistors and lead to maximal photoresponses of these transistors.[11]

The fluorination of graphene or few-layer graphene (FLG), in our case, was performed by means of treatment in an aqueous solution of hydrofluoric acid (HF).[9,12] Successful fluorination can be confirmed based on the reversible transition from a conductive to an insulating state, the high temperature stability (up to 450 °C for an activation energy of 2.0 eV), the disappearance of all peaks in the Raman spectra, a dielectric constant of 1.2 for the films from the fluorinated suspension, and the transparency of the fluorographene suspension.[9,12,13] These properties correspond to fluorographene rather than graphene oxide.[2,8] Similar processes taking place in the solution of hydrofluoric acid during the fluorination of graphene oxide have been reported.[5,14,15] The oxyfluorinated graphene layers were produced and fully characterized in terms of their chemical composition and functionalization.[15] Field-effect devices were fabricated to study the transport properties of the monolayer graphene oxyfluoride.[15] Graphene fluorination reactions in an aqueous solution of hydrofluoric acid were theoretically investigated by Lvova.[16]

---

a)E-mail: antonova@isp.nsc.ru

The interaction of F- and FHF-ions with single-crystal graphene and the grain-boundary-containing graphene surface were simulated. A grain boundary affects the interaction process; the fluorination activation energy decreases, and the heat of adsorption increases compared to defect-free graphene.

The fluorination process allows the formation of tight arrays of GQDs embedded in the fluorinated graphene network (matrix). An increase in fluorination time leads to a decrease in GQD sizes. For relatively large QDs (size ~70 nm), the quantum confinement energy levels were observed by charge spectroscopy. The relaxation time for non-equilibrium charges captured on these levels in QDs strongly depended on film thickness[10] and daylight illumination.[13] In the present study, we have focused on the processes of carrier transport in the films of fluorinated graphene with arrays of small GQDs (size < 20 nm) and on the electron structure of the fluorinated graphene part of the films. Using charge-based deep level transient spectroscopy (Q-DLTS), we have revealed the presence of a potential fluctuation inside the fluorinated part of the film and have estimated the range of these fluctuations (up to 200 meV). Comparison of these results with the properties of films with lower degrees of fluorination and higher GQD sizes[10,12] provides general insight into the electronic structure of fluorinated films with GQDs. The high potential of 2D films of fluorinated graphene with GQDs for the resistive memory effect, as well as current control in transistor structures, logic elements, and other electronic applications are also discussed in the article.

## II. SAMPLE CREATION AND CHARACTERIZATION TECHNIQUES

Pristine samples of graphene and few-layer graphene (FLG) were obtained by electrostatic exfoliation from highly oriented pyrolytic graphite (HOPG). Graphene sheets and FLG flakes ranging in thickness from monolayer to 3 nm were placed on 300 nm $SiO_2$/Si substrates. These flakes were then treated to achieve fluorination in a 3% aqueous solution of hydrofluoric acid (HF) at room temperature for different times, ranging from 0.5 to 15 min (for details, see Ref. 12). Three types of structures are compared in the present study. The less fluorinated ones are named *L-FG-GQD* (discussed in Refs. 10 and 12), and the more fluorinated films with different thicknesses are designated as *H-FG-GQD*. Further fluorination practically leads to QD disappearance; and the corresponding structures are marked as *H-FG*. The main emphasis in this study is made on the *H-FG-GQD* and *H-FG* films. Comparison with *L-FG-GQD* sheds light on the electronic structure of fluorinated films with GQDs in general.

To characterize the surface morphology in pristine and fluorinated samples, a Solver PRO NT-MDT scanning probe microscope was used. In particular, with the help of atomic force microscopy (AFM) measurements, we managed to evaluate the thickness of the graphene films in our samples. AFM measurements in the regime of frictional forces allow the visualization non-fluorinated and fluorinated graphene areas (for details, see Ref. 10). Scanning electron microscopy (SEM) images were taken using a JEOLJSM-7800F scanning electron microscope; the energy of the primary electrons in this microscope was 2 keV. Current-voltage (I–V) characteristics of graphene layers were measured using a Keithley picoammeter (model 6485) in the temperature range from 80 to 300 K. To enable the measurements of the current-voltage characteristics of the samples, silver alloy contacts were prepared on the sample surface.

Charge-based deep level transient spectroscopy Q-DLTS was used to study the carrier capture at the GQDs and carrier emission from the GQDs. A positive voltage used in the present study as a filling pulse caused the injection of electrons into the GQDs. Then, the electron emission from the GQDs was monitored using the Q-DLTS measurements. Q-DLTS measurements were performed using an ASEC-03 DLTS spectrometer by varying the time window $\tau_m$ while keeping the temperature unchanged. Here, $\tau_m = (t_2 - t_1)/\ln(t_2/t_1)$, where $t_1$ and $t_2$ are the times at which the Q-DLTS signal (due to relaxation of the dielectric-trapped charge $\Delta Q = Q(t_2) - Q(t_1)$) is measured after the end of the filling pulse. The time window $\tau_m$ at which the maximum change in the charge $\Delta Q$ was observed at different temperatures gives us the value of the carrier emission time (or charge relaxation time) for each considered temperature. The examined temperature ranges from 80 to 350 K. We analyze the times of carrier emission from the GQDs as a function of temperature. This analysis gives us information about the relaxation of excess charge from the quantum confinement levels of the GQDs. The details of the Q-DLTS study of the GQDs are given in Refs. 10 and 13.

## III. EXPERIMENTAL RESULTS

### A. Structural characterization of films with GQDs

Pristine samples of graphene and FLG films were treated in hydrofluoric acid to achieve fluorination. Figs. 1(a) and 1(b) demonstrate schematic images of the studied film on the $SiO_2$/Si substrate with two contacts before and after the fluorinating treatment. Both images in Figs. 1(c) and 1(d), observed by SEM and AFM, respectively, clearly demonstrate the corrugated surfaces of the fluorinated films with typical relief scale of 3–6 nm. A step-like increase in resistivity from ~500 Ω/□ to $10^{11}$ Ω/□ with an increase in treatment time was due to the formation of the insulated fluorographene network. The transition to the insulated state correlated with the formation of a corrugated surface relief with a period of ~100 nm (Figs. 1(c) and 1(d)). The main reason for corrugation was the difference in the lattice constants of graphene and fluorinated graphene.[2] The period of film corrugation (Figs. 1(c) and 1(d)) determined the GQD density of ~$4 \times 10^{10}$ cm$^{-2}$. GQD sizes were estimated by means of AFM (performed in both surface relief and lateral force modes) and SEM measurements. Such measurements provided visualization of the graphene and fluorinated graphene regions on the sample surface.[10] All these sizes were estimated from AFM images in the lateral-force mode; and size scattering was approximately 50%. The GQD sizes for *H-FG-GQD* were ~20 nm (Fig. 1(d)) with the fluorinated part of approximately ~80 nm. In the case of the *H-FG* structure, the GQDs became too small to estimate their sizes from the AFM images. Short-time fluorination treatment in aqueous solution of HF led to a higher GQD size of up to ~70 nm (*L-FG-GQD*) and to a decrease in fluorinated barrier thickness within the ~100 nm period. Selections of AFM images in lateral force mode for all used structures are given in Table I.

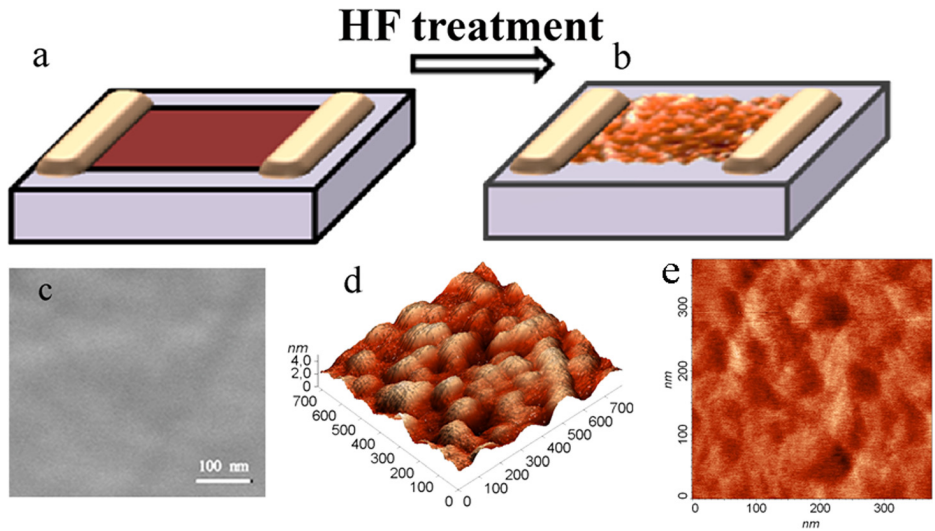

FIG. 1. Schematic view of structures with a graphene layer on SiO$_2$/Si substrate and two planar contacts before (a) and after (b) HF treatment, used for DLTS measurements. (c) Image of the surface of bilayer graphene after a 3 min treatment in an aqueous solution of HF, obtained with SEM. (d), (e) AFM measurements performed in both surface relief (corrugated surface) (d) and lateral force (e) modes for bilayer graphene films, treated in an aqueous solution of HF for 2 min (time of resistivity increase). The bright color in (e) represents fluorinated graphene, and the dark color corresponds to graphene.

### B. Carrier capture and emission on/from GQDs in high fluorinated graphene films

To characterize the fluorinated part of the films and the carrier capture and emission from the GQDs, we used charge-based deep-level transient spectroscopy. Q-DLTS spectra and Arrhenius plots for *H-FG-GQD* are shown in Fig. 2. For the Arrhenius plots, such parameters as carrier emission (or relaxation) time $\tau_m$ (the position of the DLTS peak maximum) and the temperature of the measurement are used. The activation energies for these structures with different thicknesses are small and are summarized in Table I. A relatively large energy (0.21 eV) was found only in the case of bilayer graphene. The Q-DLTS measurements were also performed in the holder, provided with an optical window. Daylight illumination with an intensity of ~$10^{17}$ photons/cm$^2$ s (Ref. 17) at the films gives practically the same energy values. The increase in peak magnitude observed for the daylight measurements (Fig. 2(a)) corresponds to a higher density of charge. The trap density N in dark conditions was estimated at $(1–2) \times 10^{11}$ cm$^{-2}$, whereas, in the case of daylight illumination, the trap density strongly depends on the temperature. At room temperature, the densities are practically the same; for lower temperatures, N is increased up to $\sim(2\text{–}5) \times 10^{11}$ cm$^{-2}$ for 80 K. The time of the non-equilibrium charge relaxation for *H-FG-GQD* does not strongly depend on the film thickness or additional illumination (Fig. 3) and equals $\sim 10$ $\mu$s. This means that some channels for carrier transport without thermal activation exist in the fluorinated films. These channels are most likely connected with carrier tunneling through potential barriers in the fluorinated part. This process is directly observed in the Q-DLTS results as the zero activation energy of the non-equilibrium carrier relaxation, which implies a complicated electronic structure (potential fluctuations) of the fluorinated potential barrier. In contrast, the time of the non-equilibrium charge relaxation for *L-FG-GQD* strongly decreases with film thickness in dark conditions. This fact is discussed in more detail below.

TABLE I. Activation energies, $E_i$, measured in the dark, and, $E_i$*, measured under daylight illumination, for carrier emission from GQDs as extracted from the Arrhenius plots. Data for L-FG-GQD films with different thicknesses (bilayer graphene, BG, and FLG with a thickness of 2 nm) in which the size of GQDs is ~70 nm and the width of fluorinated network between GQDs is ~30 nm (taken from Refs. 10 and 13). *H-FG-GQD* and *H-FG* are the same structures, additionally treated for further fluorination; in the case of *H-FG-GQD*, the size of the GQDs decreases to ~20 nm and the width of fluorinated network between the GQDs is ~80 nm; in the case of *H-FG*, the size of the GQDs becomes smaller than 10 nm. AFM images correspond to FLG films, measured in the regime of lateral forces.

| Structure | Condition of measurements | Film thickness BG | Film thickness FLG 2 nm | AFM LF image |
|---|---|---|---|---|
| L-FG-GQD | Dark | E1 = 0.09 eV<br>E2 = 0.18 eV<br>E3 = 0.33 eV | E1 = 0.12 eV<br><br>E3 = 0.35 eV | 200 nm |
|  | Daylight | E1* = 0.00 eV<br>E2* = 0.07 eV<br>E3* = 0.16 eV | E1* = 0.00 eV<br>E2* = 0.03 eV<br>E3* = 0.12 eV |  |
| H-FG-GQD | Dark | E1 = 0.00 eV<br>E2 = 0.03 eV<br>E3 = 0.21 eV | E1 = 0.00 eV<br>E2 = 0.01 eV<br>E3 = 0.05 eV | 200 nm |
|  | Daylight | E1* = 0.00 eV<br>E2* = 0.04 eV<br>E3* = 0.20 eV | E1* = 0.00 eV<br>E2* = 0.01 eV<br>E3* = 0.05 eV |  |
| H-FG | Dark |  | E1 = 0.00 eV<br>E2 = 0.04 eV<br>E3 = 0.07 eV | 200 nm |
|  | Daylight |  | E1* = 0.00 eV<br>E2* = 0.04 eV<br>E3* = 0.06 eV |  |

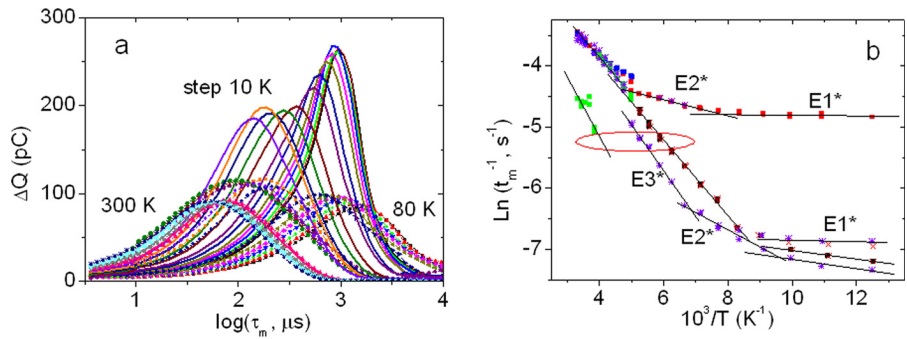

FIG. 2. (a) Q-DLTS spectra measured on a few-layer graphene film with GQDs, treated in an aqueous solution of HF for 5 min (H-FG-GQD structure) in the closed holder (points) and in the holder with a window for daylight (lines). For temperatures 300–280 K, the dark and the daylight measurements give nearly identical curves. The constant voltage during measurements is 0 V, and the magnitude of the filling pulse applied to the sample is ±12 V. (b) Arrhenius plots for Q-DLTS spectra from the sample described in (a) measured at different values of the filling pulse, +12 V and −12 V.

Q-DLTS spectra for the *H-FG* structures are similar to the results of *H-FG-GQD*. Activation energies are listed in Table I, and trap densities also increase with the decrease in temperature. This fact suggests that the Q-DLTS measurements are attributed to the fluorinated part of the films.

### C. Resistive switching and field effect for fluorinated graphene films with GQDs

Understanding the electronic structure of the 2D system of fluorinated graphene with GQDs as a function of the degree of fluorination (GQD size) is very important for applications. Electrical characterization of these films demonstrates a resistive switching effect and the possibility to modulate the current by gate voltage in transistor-like structures.

Source-drain current-voltage characteristics $I_{ds}(V_{ds})$ of one of the partially fluorinated films measured at different temperatures are given in Fig. 4(a). Measurements were made for these structures when the time of the fluorinating treatment was equal to the time of transfer into the insulator state (*L-FG-GQD*). The unipolar resistive switching effect is clearly observed in the temperature range 100–300 K. There is no appreciable hysteresis or switching effects at the current curves for negative voltages. According to the classification given in Yang's review,[18] the observed effect is nonpolar threshold switching. Measurements were performed for up to 150 cycles and demonstrated the stability of the resistive switching effect. Transistor characteristics $I_{ds}(V_g)$, measured with the use of a Si substrate as a gate for the same structure, are presented in Fig. 4(b). Two side voltage sweeps also lead to a large voltage hysteresis. This characteristic shows possible current variation ON/OFF up to $10^4$–$10^5$ times if the film with GQDs is considered as a channel in the transistor structure.

Hysteresis and unipolar resistivity switching for other structures measured after treatment in the aqueous solution of HF for 2 and 4 min are shown in Figs. 4(c) and 4(d), respectively. In the case of the low degree of fluorination (2 min treatment, corresponding to the conductive state of the film with incompletely formed GQDs, Fig. 4(c)), we observed a simple hysteresis in current-voltage curves, most likely connected with the carrier capture on the already existed GQDs. In current-voltaic characteristics of the films with a higher degree of fluorination (*H-FG-GQD*, Fig. 4(d)), a stable and reproducible resistivity switching was found at the positive voltages. One can see that an increase in degree of fluorination leads to a change in the film properties; bipolar hysteresis in current-voltage curves changed to unipolar resistivity switching.

### IV. DISCUSSION

Let us start with the results obtained by means of the Q-DLTS measurements. As it was mentioned above for *H-FG-GQD* structures, there is no dependence of the carrier emission time $\tau_m$ on film thickness. However, in the case of a relatively low degree of fluorination (structures *L-FG-GQD* analyzed in Ref. 10), the carrier emission time $\tau_m$ strongly depends on the film thickness. Fig. 5(a) shows the sum of temperature dependences of the carrier emission time $\tau_m$ for the films with different thicknesses from 1 monolayer to 3 nm for more detailed analysis. The dependencies of $\tau_m$ on film thickness $d$ for different temperatures are given in Fig. 5(b). A strong decrease in $\tau_m$ value (four orders of magnitude) exhibited clearly determining functional dependence, describing $\tau_m$ changes: $\tau_m = \tau_o exp(-d(nm)/0.35) \approx \tau_o exp(-n)$, where $n$ is the number of monolayers in the film. The physical origin of $\tau_m \sim exp(-n)$ is based on layer-by-layer fluorination. Taking into account the temperature dependence of $\tau_m$ from Ref. 10 and the energies of the GQD quantum confinement level extracted from the Q-DLTS measurements, a phenomenological expression that describes $\tau_m$ dependencies on temperature and film thickness was suggested

$$\tau_m = A \exp\left(\frac{E_a}{kT}\right)\exp\left(-\frac{d}{d_o}\right), \quad (1)$$

where $A = 8 \times 10^{-5}$ s, $d_o = 0.35$ nm, k is the Boltzmann constant, and $E_a$ is the activation energy from the quantum confinement level of the QDs, which determines the charge relaxation process. The curves in Fig. 5(a) correspond to this expression for $E_a = E_1 = 0.14$ eV and $E_a = E_2 = 0.29$ eV. The values of parameter $A$ are equal to $8.0 \times 10^{-8}$ and $8.4 \times 10^{-9}$ for graphene and bilayer graphene, respectively. Parameter $A$ for GQDs is analogous to the cross section for deep traps, i.e., any change in the vicinity of GQD causes a strong change in the value of $A$. A good correlation is clearly observed between the data, obtained from expression (1) and the experimental results.

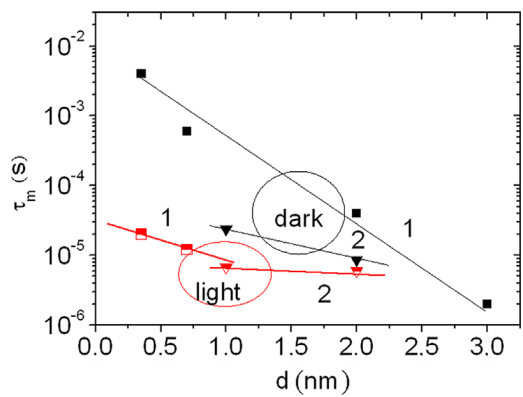

FIG. 3. The time of non-equilibrium charge relaxation $\tau_m$ as a function of film thickness for two degrees of fluorination (structures: 1 is L-FG-GQD and 2 is H-FG-GQD). "Light" and "dark" denote measurements in dark and daylight conditions. Line 1 (for dark conditions) for L-FG-GQD structures is drawn by the least squares method, and the other lines are shown for convenience.

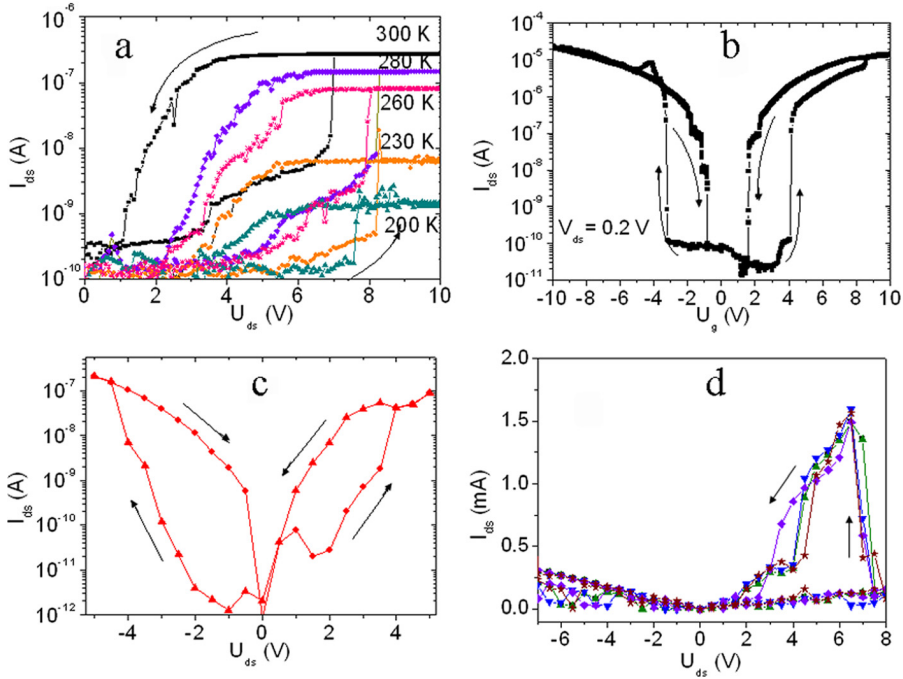

FIG. 4. (a), (b) Source-drain current measured in two directions of voltage swept as a function of (a) source-drain voltage and (b) gate voltage (the Si substrate was used as a gate). The film thickness is 1.2 nm and sample is related to the *L-FG-GQD* structure. Measurements are performed at different temperatures in (a) and at room temperature in (b). (c), (d) Source-drain characteristics measured at room temperature with two directions of voltage swept as a function of source-drain voltage. Film thickness is 2 nm, and the samples (c) and (d) are related to the *L-FG-GQD* and *H-FG-GQD* structures (2 and 4 min treatments in the aqueous solution of HF), respectively. Six cycles of I-V characteristic measurements are given in (d).

Fluorination proceeds successively, layer-by-layer, and it is apparent that the potential barrier width must decrease in the underlying monolayers. In addition, a lower degree of fluorination in the inner monolayers can be expected. Following this outline, we assume a "tapering" shape of the fluorographene barriers and a truncated pyramidal shape of GQDs in our films. Alternatively, an ever lower functionalization of the underlying monolayers leads to a decrease in the charge relaxation time with an increase in film thickness due to the lower barrier height or the change in the barrier structure (the appearance of potential fluctuations). Generally, not only the energy height of the potential barrier between GQDs and FG matrix but also the electronic structure of the FG can be significantly changed for dilute FG.

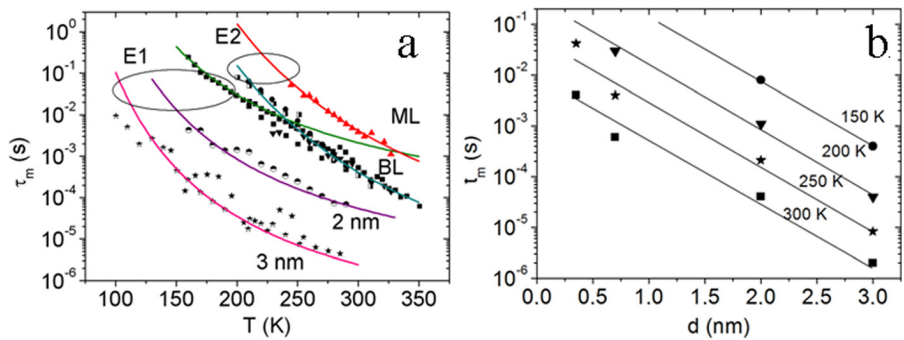

FIG. 5. (a) Emission time of charge carriers $\tau_m$ (non-equilibrium charge relaxation) as a function of temperature for the *L-FG-GQD* structures from Ref. 10. The thickness of the measured films is given as a parameter (ML and BL denote a monolayer and bilayer of graphene, respectively). The curves correspond to expression (1) for values of $E_1 = 0.14$ eV and $E_2 = 0.29$ eV. (b) Emission time of the charge carriers as a function of the layer thickness for different temperatures. The lines correspond to the equation $\tau_m = \tau_o exp(-d[nm]/0.35)$.

A strong decrease of the activation energies for the *H-FG-GQD* and *H-FG* structures (see Fig. 2 and Table I) and the appearance of a non-activated process in the charge relaxation phenomenon distinguish these structures from *L-FG-GQD*. In the case of *L-FG-GQD*, Q-DLTS measurements under illumination demonstrate a strong decrease in the activation energies, whereas, for *H-FG-GQD* structures, daylight illumination does not lead to any pronounced changes in the activation energies. An increase in the fluorinated part and a decrease in GQD size lead to the situation in which carrier migration inside the fluorinated part becomes the main process during charge relaxation. This finding is very important for understanding the carrier transport in 2D systems with GQDs and is presented in the band diagrams, given in Figs. 6(a) and 6(b) for *L-FG-GQD* and *H-FG-GQD* structures.

The low values of the relatively small activation energies are assumed to correspond to the case of charge carriers overcoming the low-height potential fluctuation occurring along the migration paths of these carriers in the fluorinated graphene matrix. The appearance of these fluctuations is caused by non-uniform fluorination. Unchanged values of activation energies obtained in dark and daylight conditions most likely result from the fact that the carrier migration path is rather long in cycling excitations/thermalization of carriers; as a result, thermally activated carrier migration is the main process inside the multibarrier fluorinated part. The increase in carrier density involved in the carrier relaxation processes due to illumination confirms this tentative scenario. The relatively large fluorinated graphene barrier heights determined for the *L-FG-GQD* structures and used for calculating the curves in Figs. 5(a) and 5(b) from expression 1 are most likely connected with carrier activation from different quantum confinement levels in relatively large GQDs. The decrease in GQD size in the *H-FG-GQD* structures leads to the situation in which activation from quantum confinement levels in relatively small GQDs is indistinguishable from carrier migration in the fluorinated part with potential fluctuations.

GQDs are known to have a wide spectrum of applications.[19–21] According to Fig. 4, two areas of possible applications are clearly revealed for fluorinated films with QDs. First of all, fluorinated graphene is a candidate material for resistive memory devices. Unipolar threshold resistive switching seems to be stable for at least up to 150 cycles (measurements were not performed for more cycles). The relation of current in different resistivity states is high ($10^2$–$10^4$). The physical reason for the unipolar switching effect is suggested to be the change in effective contact area and the nanoelectromechanical field effect[22,23] due to the corrugated surface of the fluorinated films. The switching voltage is found to depend on the degree of fluorination; the unipolar switching effect was observed only in the case when corrugation of the film occurs. Possible use of the nanoelectromechanical field effect for the creation of the Nonvolatile Random Access Memory was demonstrated, for instance, for carbon nanotubes in Ref. 24. Of course, here we have demonstrated only the possibility of observing the resistive switching effect for 2D films with GQDs. The bipolar resistivity switching effect due to the introduction of electrically active traps in the fluorinated graphene band gap under applied voltage was recently revealed in films created from a fluorinated graphene suspension.[25] In this case, we have non-corrugated films and another type of resistivity effect. It means that the unipolar switching effect observed in the present study can have a more complicated origin than the nanoelectromechanical field effect mentioned above.

The second type of possible applications is based on current modulation in the range of $I_{on}/I_{off} \sim 10^3$–$10^4$ (where $I_{on}$ and $I_{off}$ are currents in open and closed states of the transistor) at the initial stage of GQD formation before conductivity blocking by the fluorographene network (Fig. 4(b)). In this case, current modulation is combined with relatively high carrier mobility. For transistor structures with few-layer graphene channel displaced on the $SiO_2$/Si substrate, the carrier mobility is as high as 2800 cm$^2$/V s. After the formation of the fluorinated graphene network (the case considered in the present study), the current range in nano- and microamperes is observed with an increased ratio $I_{on}/I_{off} \sim 10^4$–$10^5$. The physical origin of current modulation is based on the formation of tunable potential barriers (fluorinated barriers) in the transistor channel. Multibarriers in the 2D system of fluorinated graphene with GQDs make these films promising for such applications as transistors, logic elements, etc., which require strong modulation of current.

Another property of the fluorinated graphene films with GQDs that is very important for applications is their stability. The approach used for the creation of this material is based on the chemical functionalization of graphene. Repeated Q-DLTS and current–voltage measurements demonstrated good reproducibility of the film characteristics for more than six months. The formation of defects or dangling bonds is not observed in the films during functionalization, in contrast to the case of nanostructuring. This statement is supported by our Q-DLTS measurements; the deep level electrically active defects are not found in films with GQDs. As a result, the fluorinated graphene films with GQDs have higher and longer-term chemical stability in comparison with QDs formed from suspensions.[26,27] This enhanced stability is essential for all possible applications.

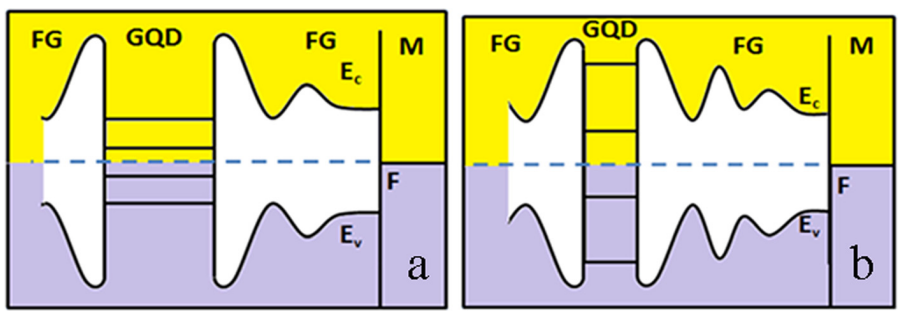

FIG. 6. (a) Sketch of band diagrams for (a) the *L-FG-GQD* structures with relatively large QDs and (b) the *H-FG-GQD* structures with small QDs in the matrix of fluorinated graphene.

Generally, the non-uniform electronic structure of the potential barrier of fluorinated graphene explains the main findings concerning carrier transport in films with GQDs. The potential fluctuation provides the possibility of carrier exchange between GQDs and carrier transport in the films instead of complete blocking of the GQDs. This is the property of the films with arrays of GQDs that provides significant potential for different electronic applications, such as nonvolatile resistive memory devices and 2D connections with optical control and logic elements.

## V. CONCLUSIONS

2D films of fluorinated graphene or few-layer graphene with tight arrays of graphene quantum dots, self-formed during chemical functionalization in an aqueous solution of hydrofluoric acid, are considered in the present study. Systems of activation energies (levels) attributed to transport and/or recharging of GQDs are compared for structures with different relations between GQD size and the fluorographene parts. One of the main findings of the present study is the high fluctuations of potential, on a scale of up to $\sim$0.2 eV, within the fluorographene parts that are due to non-uniform fluorination. The non-uniformity of the potential provides the possibility of carrier exchange between the QDs and carrier transport in the films instead of complete blocking of the GQDs. The empirical relation for the dependence of non-equilibrium carrier relaxation time on the thickness of the film with GQDs (exponential decrease with increase in the film thickness) is suggested for the case in which carrier transport is determined by carrier capture on GQDs. The physical origin of this dependence is based on the decrease in the effective potential barrier height and/or width due to the layer-by-layer fluorination of few-layer graphene. This effect allows control of the carrier emission time from GQDs by varying the layer thickness. The low values of emission time are very promising for applying graphene GQDs in creating devices with high operational speed. Daylight illumination affects the charge relaxation time only in the case of relatively large GQDs. When transport is determined by carrier migration inside the areas of the fluorinated graphene between the GQDs, the effect of illumination vanishes. The unipolar resistive switching effect and strong current modulation in transistor-like structures have been revealed in fluorinated graphene films with GQDs. The chemical and long term stability of 2D films of fluorinated graphene with GQDs is an important property for applications.

## ACKNOWLEDGMENTS

This study was partially funded by the Russian Foundation for Basic Research (Grant No. 15-02-02189).